\documentstyle[12pt]{article}

\newcommand{\Dir}{\kern -6.4pt\Big{/}}%su lettere italiane minuscole
\newcommand{\Dirin}{\kern -10.4pt\Big{/}\kern 4.4pt}
\newcommand{\DDir}{\kern -7.6pt\Big{/}}%su lettere italiane maiuscole
\newcommand{\DGir}{\kern -6.0pt\Big{/}}%su lettere greche
\newcommand{\sla}{\kern -5.4pt /}
\newcommand{\ar}{\rightarrow}

\newcommand{\be}{\begin{equation}}
\newcommand{\ee}{\end{equation}}
\newcommand{\bea}{\begin{eqnarray}}
\newcommand{\eea}{\end{eqnarray}}
\newcommand{\beanon}{\begin{eqnarray*}}
\newcommand{\eeanon}{\end{eqnarray*}}
\newcommand{\ba}{\begin{array}}
\newcommand{\ea}{\end{array}}
\newcommand{\bi}{\begin{itemize}}
\newcommand{\ei}{\end{itemize}}
\newcommand{\ben}{\begin{enumerate}}
\newcommand{\een}{\end{enumerate}}
\newcommand{\bc}{\begin{center}}
\newcommand{\ec}{\end{center}}

\newcommand{\ol}{\overline}
\newcommand{\dotp}{\!\cdot\!}

\newcommand{\NP}[1]{{\it Nucl.\ Phys.\ }{\bf #1}}
\newcommand{\PL}[1]{{\it Phys.\ Lett.\ }{\bf #1}}

\newcommand{\PR}[1]{{\it Phys.\ Rev.\ }{\bf #1}}

\newcommand{\HPA}[1]{{\it Helv.\ Phys.\ Acta.\ }{\bf #1}}

\begin{document}
\baselineskip .6 truecm

\pagestyle{empty}
\setcounter{page}{0}

\begin{flushright}
{\large DFTT 14/94}\\
{\rm April 1994\hspace*{.5 truecm}}\\
\end{flushright}

\vspace*{\fill}

\begin{center}
{\Large \bf Heavy Quark Production at $e^+ e^-$ Colliders
in Multijet Events and a New Method of Computing
Helicity Amplitudes\footnote{ Talk presented at the XXIXth Rencontres de
Moriond,
QCD and High Energy Hadronic Interactions, \hfill\break\indent
\hskip .3 truecm M\'eribel - Savoie - France \ March 19--26, 1994.
\hfill\break\indent
\hskip .3 truecm  e--mail: ballestrero,maina,moretti@to.infn.it}}\\[2cm]
{\large Alessandro Ballestrero, Ezio Maina and Stefano Moretti}\\[.3 cm]
{\it Dipartimento di Fisica Teorica, Universit\`a di Torino}\\
{\it and INFN, Sezione di Torino}\\
{\it v. Giuria 1, 10125 Torino, Italy.}
\end{center}

\vspace*{\fill}

\begin{abstract}
{\normalsize
Heavy quark production in multijet events at $e^+e^-$ colliders
is studied at tree level. Total production rates
are given and compared with the corresponding results
for massless quarks. A new method of computing helicity amplitudes is
briefly sketched.
}
\end{abstract}

\vspace*{\fill}

\newpage

\subsection*{Introduction}
The great number of hadronic decays of the $Z^0$ observed at LEP
provides the opportunity to test our understanding of strong interactions
in unprecedented detail.
Large samples of multi--jet events
have been accumulated and analyzed \cite{expMC,alphas}.
Recent advances in $b$--tagging techniques based on the introduction of
vertex detectors and on a refinement of the selection procedures,
with their large efficiencies and the resulting high purities, have paved
the way to the study of heavy--quark production in association with
light--quark and gluon jets. \par
It has been pointed out in series of papers \cite{last1,last2,last3}
that the effects of the $b$--quark mass are substantial
and increase with the number of jets.
We have studied jet production at tree
level taking full account of $\gamma, Z$ interference
and of quark masses.
These latter reduce the available phase space and strongly decrease the
emission of gluons collinear with the quark direction, the so called ``dead
cone effect''.
In particular we have computed:
\begin{description}
\item[] $\bullet$ $ e^+e^- \ar Q \bar Q g $ \hfill
 $\bullet$ $e^+e^- \ar Q \bar Q g g$ \hfill
$\bullet$ $ e^+e^- \ar Q_1 \bar {Q_1} Q_2 \bar {Q_2}  $
\item[]$\bullet$ $ e^+e^- \ar Q \bar Q \gamma $ \hfill \mbox{\hspace*{-.15
truecm}}
$\bullet$ $ e^+e^- \ar Q \bar Q \gamma g $ \hfill
$\bullet$ $ e^+e^- \ar Q \bar Q \gamma \gamma $ \mbox{\hspace*{.7 truecm}}
\item[] \hfill $\bullet$ $ e^+e^- \ar Q \bar Q g g g $ \hfill
$\bullet$ $ e^+e^- \ar Q_1 \bar {Q_1} Q_2 \bar {Q_2} g $ \hfill \hspace*{.1
truecm}
\end{description}
We have used  $M_Z=91.1$ GeV, $\Gamma_Z=2.5$ GeV,
$\sin^2 (\theta_W)=.23$, $m_b=5.$ GeV, $\alpha_{em}= 1/128$ and
$\alpha_{s}= .115$ in the numerical part of our work.\par
When the number of Feynman diagrams becomes large it is convenient to
compute the amplitudes using helicity methods instead of computing
directly the amplitude squared. We have used two of the most popular formalism
\cite{hz,ks} in our calculations.
Both methods can be easily implemented in a small set of nested subroutines.
This however results in  computer programs which are too slow.
For the five--jet case we have resorted to the symbolic package $Mathematica$
\cite{math}
to write down the Fortran expression for each helicity amplitude.
With this procedure we have produced a rather large piece of code,
which however runs quite fast, and therefeore
can be used in high statistics Montecarlo runs.
As an example the program for $q\bar q ggg$ production is about
24,000 lines long, but requires only about $5\times 10^{-2}$ seconds
to evaluate on a Vaxstation 4000/90. This is still acceptable but clearly
indicates that faster methods are needed, as the one we have recently
developed and used for computing $e^+e^- \ar b \bar b W^+ W^-$
\cite{method,next}.

\subsection*{A new method for helicity amplitude calculations}
There are normally three possible ways of evaluating a spinor line.
The first consists in reducing  the expression to
a trace \cite{cr}.
The second amounts to writing
explicitely, for example in the helicity representation \cite{hz},
the components of the spinors, of the vertex matrices $\eta\sla_i$'s
and of the $p\sla_i$
and then proceed to the multiplication of the matrices and spinors.
The other way \cite{calkul,ks} consists in decomposing every $p\sla_i$ in
sums of external momenta $k\sla_i$ and use  the relation
$k\sla=\sum_{\lambda}U(k,\lambda)\ol U(k,\lambda)+M$ (with $M=+m$ if
$U=u$, $M=-m$ if $U=v$) in order to reduce everything to the computation
 of expressions of the type $\ol U(k_i,\lambda_i)\eta\sla U(k_j,\lambda_j)$.
\par
We get a remarkable simplification with respect to the procedures sketched
above
inserting just before every $(p\sla _i+\mu_i)$ in a spinor line,
completeness relations formed with eigenvectors of $p\sla _i$.
To do this we  must  construct  spinors $U(p,\lambda)$  which  are
defined
also for $p$ spacelike. With this method, in addition to
reducing ourselves to the computation of expressions of the type
$\ol U(p_i,\lambda_i) \eta\sla_i U(p_j,\lambda_j)$, we avoid the
proliferation  of terms due to the decomposition of the  $p\sla_i$
in terms of external momenta.\par

One can easily costruct an example of spinors defined
for any value of $p^2$   and   satisfying   Dirac   equation   and
completeness relation,
with a straightforward generalization of those introduced in ref.\cite{ks}.
One first defines spinors $w(k_0,\lambda)$ for an auxiliary
massless vector $k_0$ satisfying
\be
w(k_0,\lambda)\bar{w}(k_0,\lambda)=\frac{1+\lambda\gamma_5}{2}k\sla_0
\ee
and with their relative phase fixed by
\be
w(k_0,\lambda)=\lambda k\sla_1 w(k_0,-\lambda),
\ee
with $k_1$  a second auxiliary vector such that $k_1^2=-1$,
$k_0\dotp k_1=0$.
Spinors for a  four momentum $p$, with
$m^2=p^2$ are then obtained as:
\be\label{uvks}
u(p,\lambda)=\frac{p\sla + m}{\sqrt{2\,p\dotp k_0}}\;w(k_0,-\lambda)
\hskip 1 truecm
v(p,\lambda)=\frac{p\sla - m}{\sqrt{2\,p \dotp k_0}}\;w(k_0,-\lambda)
\ee
\be\label{uvksconj}
\bar u(p,\lambda)=\bar w(k_0,-\lambda)\;\frac{p\sla + m}{\sqrt{2\,p \dotp k_0}}
\hskip 1 truecm
\bar v(p,\lambda)=\bar w(k_0,-\lambda)\;\frac{p\sla - m}{\sqrt{2\,p \dotp k_0}}
\ee
If $p$ is spacelike,  one of the two determination of $\sqrt{p^2}$
has  to  be chosen for $m$ in the  above  formulae,  but  physical
results will not depend on this choice.

One can readily  check that with the previous definitions,
Dirac equations
\be\label{dirac}
p\sla u(p)=+m u(p) \hskip 2 truecm
p\sla v(p)=-m v(p)
\ee
\be\label{diracconj}
\bar u(p) p\sla =+m \bar u(p) \hskip 2 truecm
\bar v(p) p\sla =-m \bar v(p)
\ee
and the completeness relation
\be\label{compl}
1=\sum_\lambda\frac{u(p,\lambda)\bar{u}(p,\lambda)-v(p,\lambda)
\bar{v}(p,\lambda)}{2m}
\ee
are satisfied also when $p^2\leq 0$ and $m$ is imaginary.
\par
Let us now consider the case in which there are only two
insertions in a spinor line:
\be\label{T2}
T^{(2)}(p_1;\eta_1;p_2;\eta_2;p_3)=\ol{U}(p_1,\lambda_1)\eta\sla_1(p\sla_2+
\mu_2)\eta\sla_2 U(p_3,\lambda_3).
\ee
One can insert in eq. (\ref{T2}), on the left of $(p\sla_2+\mu_2)$
the relation (\ref{compl}) and make use of Dirac equations to get:
\bea
T^{(2)} & = &\frac{1}{2}\ol{U}(p_1,\lambda_1)\eta\sla_1 u(p_2,\lambda_2)\times
\bar{u}(p_2,\lambda_2)\eta\sla_2 U(p_3,\lambda_3)\times \left( 1+{\mu_2\over
m_2}\right)+\nonumber \\ \label{T2uv}
    &   &
\frac{1}{2}\ol{U}(p_1,\lambda_1)\eta\sla_1 v(p_2,\lambda_2)\times
\bar{v}(p_2,\lambda_2)\eta\sla_2 U(p_3,\lambda_3)\times \left( 1-{\mu_2\over
m_2}\right)
\eea
This example can  be generalized to any number of insertions and shows
that the factors $(p\sla_i+\mu_i)$ can be eliminated,
reducing all fermion lines essentially to  products of $T$ functions:
\be\label{T}
T_{\lambda_1 \lambda_2}(p_1;\eta;p_2)=\ol{U}(p_1,\lambda_1) \eta\sla U(p_2,
\lambda_2)
\ee
defined for any value of $p_1^2$ and $p_2^2$.

The $T$ functions (\ref{T}) have a simple dependence on $m_1$
and $m_2$:
\be\label{Texpr}
\widetilde T_{\lambda_1 \lambda_2}(p_1;\eta;p_2)\equiv
\sqrt{p_1\dotp k_0}\;\sqrt{p_2\dotp k_0}\;
T_{\lambda_1 \lambda_2}(p_1;\eta;p_2)=\hskip 4truecm
\ee
\[
A_{\lambda_1 \lambda_2}(p_1;\eta;p_2)
+M_1B_{\lambda_1 \lambda_2}(p_1;\eta;p_2)
 +M_2C_{\lambda_1 \lambda_2}(p_1;\eta;p_2)
+M_1M_2D_{\lambda_1 \lambda_2}(p_1;\eta;p_2)\nonumber
\]
where
\be\label{UM}
M_i=+m_i \hskip 3truemm \mbox{if}\hskip 3truemm
U(p_i,\lambda_i)=u(p_i,\lambda_i)  \hskip 6truemm
M_i=-m_i \hskip 3truemm  \mbox
{if}\hskip 3truemm
U(p_i,\lambda_i)=v(p_i,\lambda_i).
\ee
The functions $A$, $B$, $C$, $D$ are independent of
$m_1$ and $m_2$ and of the $u$ or $v$ nature of $\ol{U}(p_1,\lambda_1)$
and $U(p_2,\lambda_2)$. Every $T^{(n)}$, for any number of insertions
turns out to be of the form (\ref{T}) and this greatly simplyfies
the rules for evaluating spinor lines \cite{method}.

\par
\subsection*{Results}
A selection of our results is shown in fig.1 through 3, and we refer
to the original papers for more details.
In fig.1b we present the cross sections for $e^+e^-\rightarrow q \bar q g$
and $e^+e^-\rightarrow q \bar q gg$ with
$q =  d, b$ as a function of $y_{\rm cut}$ for both the JADE \cite{jade}
and DURHAM \cite{durham}
definitions of $y$ at LEP I.  The ratio of massive to masslees cross section
is shown in fig.1a.
For small $y^D_{\rm cut}$ the cross
section for $b\bar b g$ is almost 20\% smaller than for $d \bar d g$.
As expected the ratio becomes closer to one for larger $y_{\rm cut}$,
but for $y_{\rm cut}$ as large as .2,
still $R^{bd}_3 = \sigma (b \bar b  g)/\sigma (d \bar d g )\leq .96$
in both schemes.\par
Jet--shape variables have been extensively studied as a tool to determine
$\alpha_s$ and as a testing ground for the agreement between data and the
standard description of strong interactions. In the ranges
used for measuring $\alpha_s$, the ratio of massive to
massless tree--level predictions can significantly differ from unity
and it depends both from the variable and from its actual value.
We have compared at $O(\alpha_s)$
the ratio $R_{\gamma Z}$, which is obtained from the full matrix element,
with $R_{\gamma}$, the ratio which results neglecting the $Z^0$
(as in JETSET),
for Thrust, Oblateness, C-parameter, $M_H$ and $M_D$.
The difference between $R_{\gamma Z}$ and $R_{\gamma}$ turns
out to be about $1.2\times 10^{-2}$, almost independent of the particular
variable and of its specific value.
As an example, in fig.2 we show both $R_{\gamma Z}$  and $R_{\gamma}$ for the
$M_H$, $M_D$, $C$ and $O$ distributions.\par
In fig.3 we show the ratios $\sigma(2b3g)/\sigma(2d3g)$
(continuous line), \allowbreak $\sigma(2u2bg)/\sigma(2u2dg)$ (dashed line)
and $\sigma(2d2bg)/\sigma(2d2sg)$ (dotted line) in the two
recombination schemes as a function of $y_{cut}$.
These curves confirm our previous conclusions
that mass effects increase with the number of final state light partons.
The ratio for the dominant $2q3g$ production process is equal to
.58 at $y_{cut}=.001$ in the
DURHAM scheme and to .67 at $y_{cut}=.005$ in the JADE scheme.
It is even smaller for the processes with four quark jets in the final state.
This corresponds to a 6$\div$8\% decrease in the predictions for the total
five--jet cross section.\par
When our results for jet production are compared with the data
it is to be remebered that
we have used $\alpha_s =.115$ which corresponds to
$Q^2= M^2_{z^0}$ with $\Lambda_{\overline{MS}}=200$ MeV with five
active flavours. The analysis of shape variables and jet rates to
${\cal O}(\alpha_s^2)$
has shown that, in order to get agreement between the data and the
theoretical predictions, a rather small scale for the strong
coupling constant has to be chosen \cite{alphas}.
\par
The flavour independence of the strong coupling
constant has been investigated by several groups \cite{alphab}
using both jet--rates and shape variables. Mass corrections
have played an important r\^{o}le in this study.
We look forward to more detailed analyses of QCD in the heavy
quark sector.\par

\newpage

\subsection*{Figure Captions}

\begin{description}

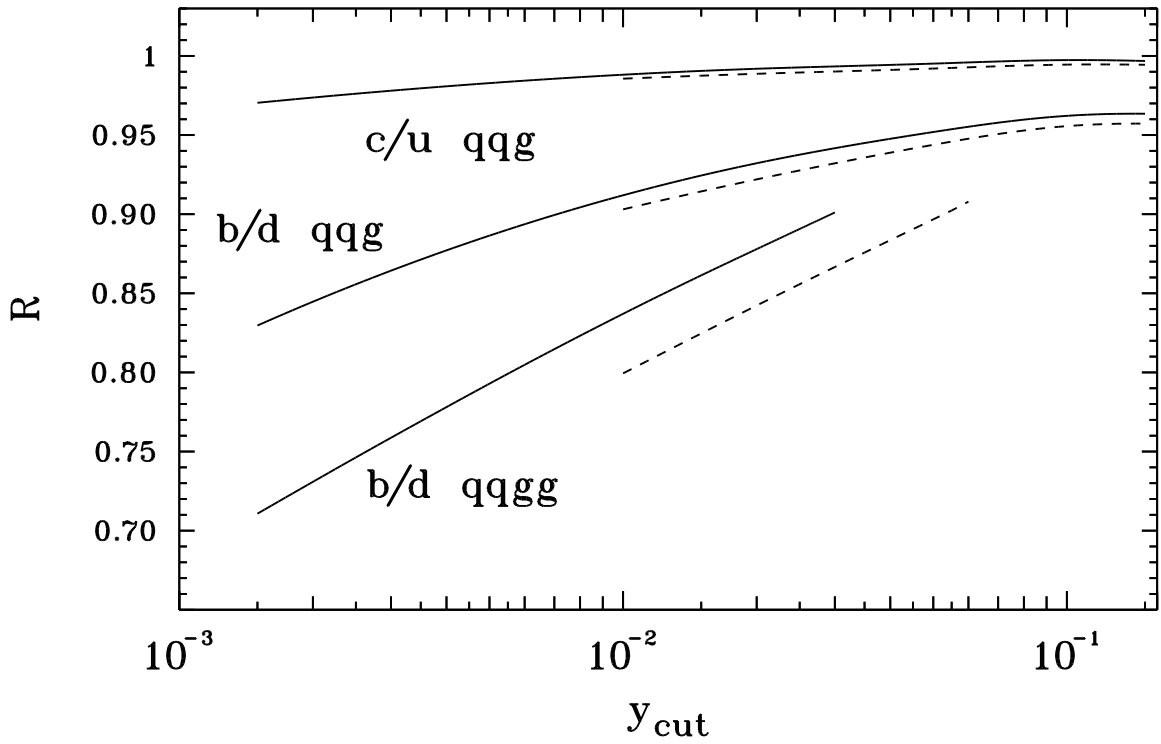
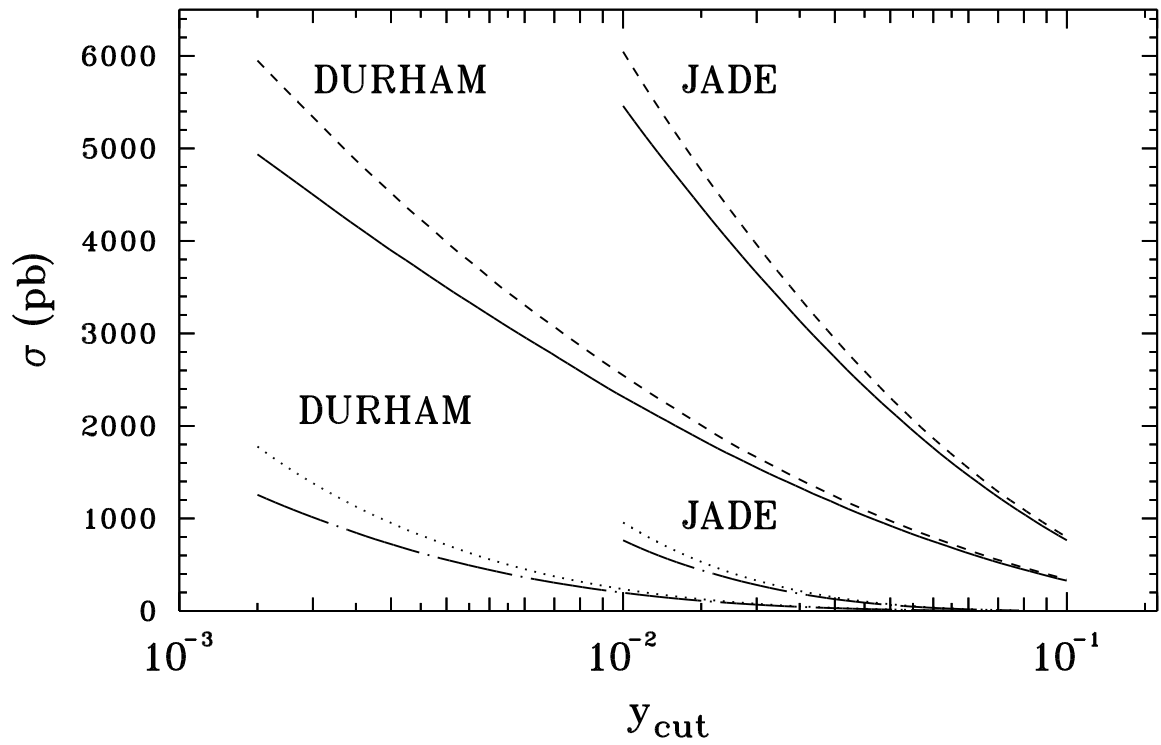
\item[Fig. 1] In the upper part  we show
$R^{cu}_3 = \sigma (c \bar c g)/\sigma (u \bar u g)$,
$R^{bd}_3 = \sigma (b \bar b g)/\sigma (d \bar d g)$ and
$R^{bd}_4 = \sigma (b \bar b g g)/\sigma (d \bar d g g)$ as a function of
$y_{\rm cut}$
for $y^J$ (dashed) and $y^D$ (continuous).
In the lower part we present the cross sections for $e^+e^-\rightarrow b \bar b
g$
(continuous), $e^+e^-\rightarrow d \bar d g$ (dashed),
$e^+e^-\rightarrow b \bar b gg$ (chain-dotted) and
$e^+e^-\rightarrow d \bar d gg$ (dotted) as a function of $y_{\rm cut}$
for both definitions of $y$ at $\sqrt{s} = 91.1$ GeV.

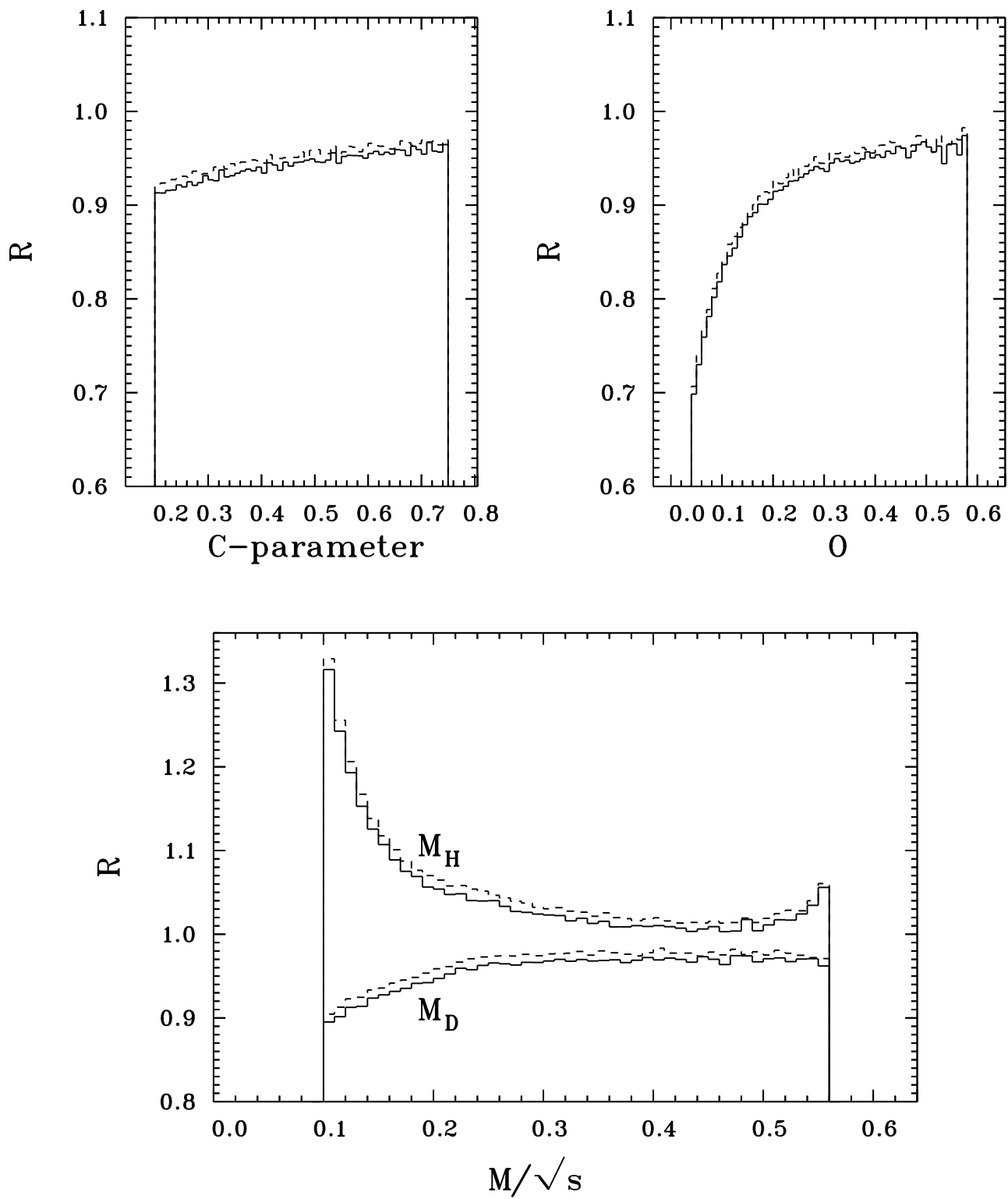
\item[Fig. 2] The ratios $ R = d\sigma (b \bar b g) / d \hat M \big{/}
                          d\sigma (d \bar d g) / d \hat M$
for $\hat M = M_H/\sqrt{s}$ and $\hat M = M_D/\sqrt{s}$ (lower part)
and for $\hat M = C-parameter$ and $\hat M = Oblateness$ (upper part)
from the full matrix element (continuous) and from the
photon contribution alone (dashed) at $\sqrt{s} = 91.1$ GeV.

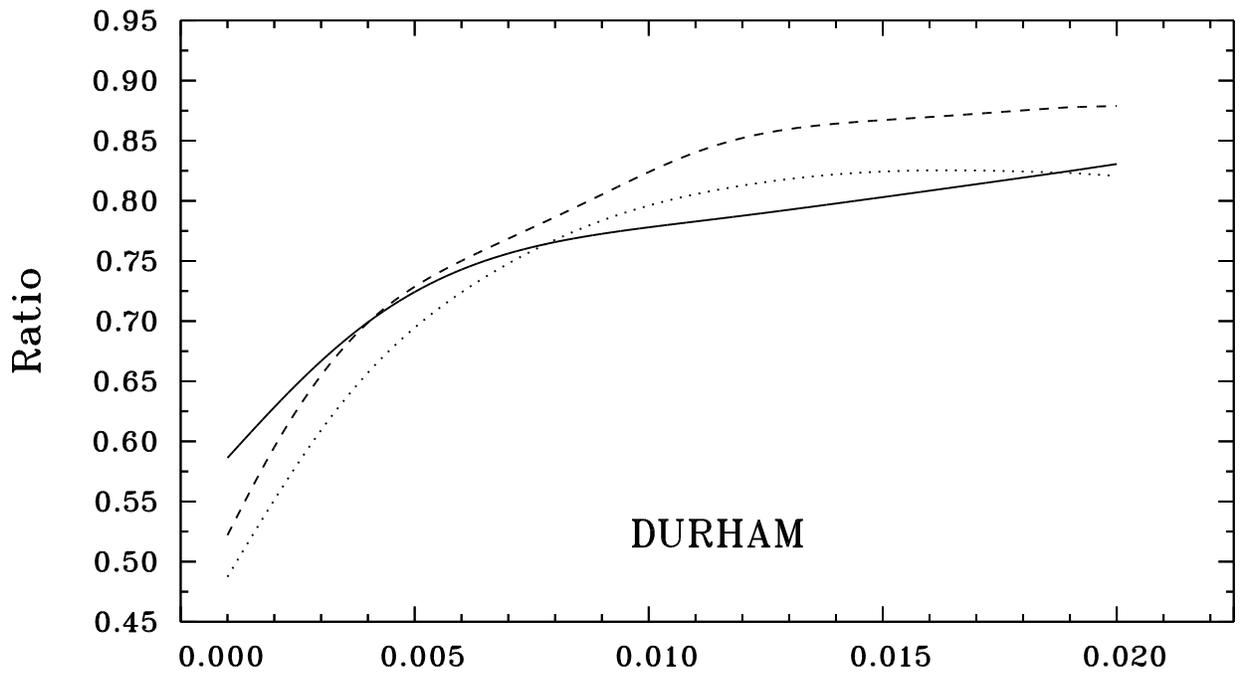
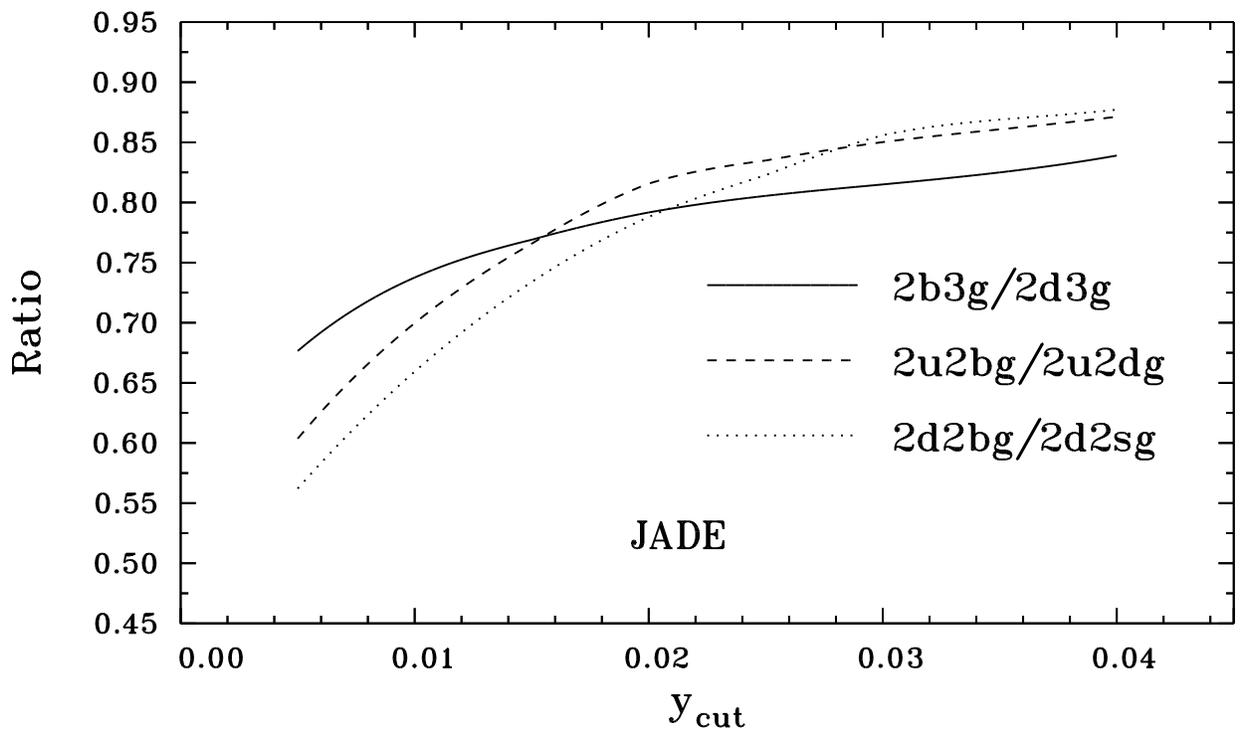
\item[Fig. 3] Ratio of massive to massless cross sections for
  $\sigma(2b3g)/\sigma(2d3g)$
(continuous line), $\sigma(2u2bg)/\sigma(2u2dg)$ (dashed line)
and $\sigma(2d2bg)/\sigma(2d2sg)$ (dotted line) in the JADE and DURHAM
recombination schemes as a function of $y_{cut}$ at $\sqrt{s} = 91.1$ GeV.

\end{description}


\baselineskip .5 truecm

\begin{thebibliography}{1}

\bibitem{expMC}
L3 Collaboration, B. Adeva {\it et al.},
Z. Phys. {\bf C 55} (1992) 39.\\
ALEPH  Collaboration, D. Buskulic {\it et al.},
Z. Phys. {\bf C 55} (1992) 209.

\bibitem{alphas}
DELPHI Collaboration, P. Abreu {\it et al.},
Z. Phys. {\bf C 54} (1992) 55.\\
OPAL Collaboration, P.D. Acton {\it et al.},
Z. Phys. {\bf C 55} (1992) 1.

\bibitem{last1}
A. Ballestrero, E. Maina and S. Moretti, Phys. Lett.
{\bf B294} (1992) 425.

\bibitem{last2}
A. Ballestrero, E. Maina and S. Moretti, Torino Preprint
DFTT 53-92, October 1992. To appear in Nucl. Phys. B.

\bibitem{last3}
A. Ballestrero and E. Maina, Phys. Lett.
{\bf B323} (1994) 53.

\bibitem{hz} K.~Hagiwara and D.~Zeppenfeld,
\NP{B274} (1986) 1.

\bibitem{ks} F.A.~Berends, P.H.~Daverveldt and R.~Kleiss
\NP{B253} (1985) 441, R.~Kleiss and W.J.~Stirling,
\NP{B262} (1985) 235.\\
C. Mana and M. Martinez, Nucl. Phys. {\bf B287} (1987) 601.

\bibitem{math}
$Mathematica$ is a registered trademark of Wolfram Research, Inc.

\bibitem{method}
A. Ballestrero and E. Maina, Torino Preprint
DFTT 76-93, January 1994.

\bibitem{next}
A. Ballestrero, E. Maina and S. Moretti, Torino Preprint
DFTT 78-93, March 1994.

\bibitem{cr}
M. Caffo and E. Remiddi, \HPA{55} (1982) 339.\\
G. Passarino, \PR{D28} (1983) 2867, \NP{B237} (1984) 249.

\bibitem{calkul}
P.~De~Causmaecker, R.~Gastmans, W.~Troosts and T.~T.~Wu,
\PL{B105} (1981) 215, \NP{B206} (1982) 53;
F.A.~Berends, R.~Kleiss,P.~De~Causmaecker, R.~Gastmans, W.~Troosts
and T.~T.~Wu,
\NP{B206} (1982) 61, {\bf B239} (1984) 382, {\bf B239} (1984) 395,
{\bf B264} (1986) 243, {\bf B264} (1986) 265.

\bibitem{jade}
JADE Collaboration, W. Bartel {\it et al.}, Z. Phys.
{\bf C 33} (1986) 23.\\
JADE Collaboration, S. Bethke {\it et al.}, Phys. Lett.
{\bf B213} (1988) 235.

\bibitem{durham}
N. Brown and W.J. Stirling, Phys. Lett. {\bf B252} (1990) 657;
Z. Phys. {\bf C 53} (1992) 629.

\bibitem{alphab}
L3 Collaboration, B. Adeva {\it et al.},
Phys. Lett. {\bf B271} (1991) 461.\\
DELPHI Collaboration, P. Abreu {\it et al.},
Phys. Lett. {\bf B307} (1993) 221.\\
OPAL Collaboration, R. Akers {\it et al.},
Preprint CERN-PPE/93-118, July 1993.

\end{thebibliography}
\end{document}